


%
%


\def\famname{
 \textfont0=\textrm \scriptfont0=\scriptrm
 \scriptscriptfont0=\sscriptrm
 \textfont1=\textmi \scriptfont1=\scriptmi
 \scriptscriptfont1=\sscriptmi
 \textfont2=\textsy \scriptfont2=\scriptsy \scriptscriptfont2=\sscriptsy
 \textfont3=\textex \scriptfont3=\textex \scriptscriptfont3=\textex
 \textfont4=\textbf \scriptfont4=\scriptbf \scriptscriptfont4=\sscriptbf
 \skewchar\textmi='177 \skewchar\scriptmi='177
 \skewchar\sscriptmi='177
 \skewchar\textsy='60 \skewchar\scriptsy='60
 \skewchar\sscriptsy='60
 \def\rm{\fam0 \textrm} \def\bf{\fam4 \textbf}}
\def\sca#1{scaled\magstep#1} \def\scah{scaled\magstephalf} 
\def\twelvepoint{
 \font\textrm=cmr12 \font\scriptrm=cmr8 \font\sscriptrm=cmr6
 \font\textmi=cmmi12 \font\scriptmi=cmmi8 \font\sscriptmi=cmmi6 
 \font\textsy=cmsy10 \sca1 \font\scriptsy=cmsy8
 \font\sscriptsy=cmsy6
 \font\textex=cmex10 \sca1
 \font\textbf=cmbx12 \font\scriptbf=cmbx8 \font\sscriptbf=cmbx6
 \font\it=cmti12
 \font\sectfont=cmbx12 \sca1
 \font\refrm=cmr10 \scah \font\refit=cmti10 \scah
 \font\refbf=cmbx10 \scah
 \def\twelverm{\textrm} \def\twelveit{\it} \def\twelvebf{\textbf}
 \famname \textrm 
 \voffset=.04in \hoffset=.21in
 \normalbaselineskip=18pt plus 1pt \baselineskip=\normalbaselineskip
 \parindent=21pt
 \setbox\strutbox=\hbox{\vrule height10.5pt depth4pt width0pt}}


\catcode`@=11

{\catcode`\'=\active \def'{{}^\bgroup\prim@s}}

\def\screwcount{\alloc@0\count\countdef\insc@unt}   
\def\screwdimen{\alloc@1\dimen\dimendef\insc@unt} 
\def\screwbox{\alloc@4\box\chardef\insc@unt}

\catcode`@=12


\overfullrule=0pt			
\voffset=.04in \hoffset=.21in
\vsize=9in \hsize=6in
\parskip=\medskipamount	
\lineskip=0pt				
\normalbaselineskip=18pt plus 1pt \baselineskip=\normalbaselineskip
\abovedisplayskip=1.2em plus.3em minus.9em 
\belowdisplayskip=1.2em plus.3em minus.9em	
\abovedisplayshortskip=0em plus.3em	
\belowdisplayshortskip=.7em plus.3em minus.4em	
\parindent=21pt
\setbox\strutbox=\hbox{\vrule height10.5pt depth4pt width0pt}
\def\makefootline{\baselineskip=30pt \line{\the\footline}}
\footline={\ifnum\count0=1 \hfil \else\hss\twelverm\folio\hss \fi}
\pageno=1


\def\boxit#1{\leavevmode\thinspace\hbox{\vrule\vtop{\vbox{
	\hrule\kern1pt\hbox{\vphantom{\bf/}\thinspace{\bf#1}\thinspace}}
	\kern1pt\hrule}\vrule}\thinspace}
\def\Boxit#1{\noindent\vbox{\hrule\hbox{\vrule\kern3pt\vbox{
	\advance\hsize-7pt\vskip-\parskip\kern3pt\bf#1
	\hbox{\vrule height0pt depth\dp\strutbox width0pt}
	\kern3pt}\kern3pt\vrule}\hrule}}


\def\put(#1,#2)#3{\screwdimen\unit  \unit=1in
	\vbox to0pt{\kern-#2\unit\hbox{\kern#1\unit
	\vbox{#3}}\vss}\nointerlineskip}

%
%
%
%
%
%
%

\def\\{\hfil\break}

\def\center{\leftskip=0pt plus 1fill \rightskip=\leftskip \parindent=0pt
 \def\textindent##1{\par\hangindent21pt\footrm\noindent\hskip21pt
 \llap{##1\enspace}\ignorespaces}\par}
\def\unnarrower{\leftskip=0pt \rightskip=\leftskip}
\def\thetitle#1#2#3#4#5{
 \font\titlefont=cmbx12 \sca2 \font\footrm=cmr10 \font\footit=cmti10
  \twelverm
	{\hbox to\hsize{#4 \hfill ITP-SB-#3}}\par
	\vskip.8in minus.1in {\center\baselineskip=1.44\normalbaselineskip
 {\titlefont #1}\par}{\center\baselineskip=\normalbaselineskip
 \vskip.5in minus.2in #2
	\vskip1.4in minus1.2in {\twelvebf ABSTRACT}\par}
 \vskip.1in\par
 \narrower\par#5\par\unnarrower\vskip3.5in minus2.3in\eject}
\def\paper\par#1\par#2\par#3\par#4\par#5\par{\twelvepoint
	\thetitle{#1}{#2}{#3}{#4}{#5}} 
\def\author#1#2{#1 \vskip.1in {\twelveit #2}\vskip.1in}
\def\ITP{Institute for Theoretical Physics\\
	State University of New York, Stony Brook, NY 11794-3840}


\def\sect#1\par{\par\ifdim\lastskip<\medskipamount
	\bigskip\medskip\goodbreak\else\nobreak\fi
	\noindent{\sectfont{#1}}\par\nobreak\medskip} 
\def\itemize#1 {\item{[#1]}}	
\def\vol#1 {{\refbf#1} }		 

\def\ref#1{\setbox0=\hbox{M}$\vbox to\ht0{}^{#1}$}


\def\NP #1 {{\refit Nucl. Phys.} {\refbf B{#1}} }
\def\PL #1 {{\refit Phys. Lett.} {\refbf{#1}} }
\def\PR #1 {{\refit Phys. Rev. Lett.} {\refbf{#1}} }
\def\PRD #1 {{\refit Phys. Rev.} {\refbf D{#1}} }


\hyphenation{pre-print}
\hyphenation{quan-ti-za-tion}

%
%

\def\on#1#2{{\buildrel{\mkern2.5mu#1\mkern-2.5mu}\over{#2}}}
\def\dt#1{\on{\hbox{\bf .}}{#1}}                
\def\ddt#1{\on{\hbox{\bf .\kern-1pt.}}#1}    
\def\slap#1#2{\setbox0=\hbox{$#1{#2}$}
	#2\kern-\wd0{\hbox to\wd0{\hfil$#1{/}$\hfil}}}
\def\sla#1{\mathpalette\slap{#1}}                
\def\bop#1{\setbox0=\hbox{$#1M$}\mkern1.5mu
	\vbox{\hrule height0pt depth.04\ht0
	\hbox{\vrule width.04\ht0 height.9\ht0 \kern.9\ht0
	\vrule width.04\ht0}\hrule height.04\ht0}\mkern1.5mu}
\def\bo{{\mathpalette\bop{}}}                        
\def~{\widetilde} 
\mathcode`\*="702A                  
\def\in{\relax\ifmmode\mathchar"3232\else{\refit in\/}\fi} 
\def\f#1#2{{\textstyle{#1\over#2}}}	   
\def\half{{\textstyle{1\over{\raise.1ex\hbox{$\scriptstyle{2}$}}}}}

\catcode`\^^?=13				    
\catcode128=13 \def €{\"A}                 
\catcode129=13 \def {\AA}                 
\catcode130=13 \def '{\c}           	   
\catcode131=13 \def ƒ{\'E}                   
\catcode132=13 \def "{\~N}                   
\catcode133=13 \def …{\"O}                 
\catcode134=13 \def †{\"U}                  
\catcode135=13 \def ‡{\'a}                  
\catcode136=13 \def ˆ{\`a}                   
\catcode137=13 \def ‰{\^a}                 
\catcode138=13 \def Š{\"a}                 
\catcode139=13 \def ‹{\~a}                   
\catcode140=13 \def Œ{\alpha}            
\catcode141=13 \def {\chi}                
\catcode142=13 \def Ž{\'e}                   
\catcode143=13 \def {\`e}                    
\catcode144=13 \def {\^e}                  
\catcode145=13 \def '{\"e}                
\catcode146=13 \def '{\'\i}                 
\catcode147=13 \def "{\`\i}                  
\catcode148=13 \def "{\^\i}                
\catcode149=13 \def •{\"\i}                
\catcode150=13 \def –{\~n}                  
\catcode151=13 \def —{\'o}                 
\catcode152=13 \def ˜{\`o}                  
\catcode153=13 \def ™{\^o}                
\catcode154=13 \def š{\"o}                 
\catcode155=13 \def ›{\~o}                  
\catcode156=13 \def œ{\'u}                  
\catcode157=13 \def {\`u}                  
\catcode158=13 \def ž{\^u}                
\catcode159=13 \def Ÿ{\"u}                
\catcode160=13 \def  {\tau}               
\catcode161=13 \mathchardef ¡="2203     
\catcode162=13 \def ¢{\oplus}           
\catcode163=13 \def £{\relax\ifmmode\to\else\itemize\fi} 
\catcode164=13 \def ¤{\subset}	  
\catcode165=13 \def ¥{\infty}           
\catcode166=13 \def ¦{\mp}                
\catcode167=13 \def §{\sigma}           
\catcode168=13 \def ¨{\rho}               
\catcode169=13 \def ©{\gamma}         
\catcode170=13 \def ª{\leftrightarrow} 
\catcode171=13 \def «{\relax\ifmmode\acute\else\expandafter\'\fi}
\catcode172=13 \def ¬{\relax\ifmmode\expandafter\ddt\else\expandafter\"\fi}
\catcode173=13 \def ­{\equiv}            
\catcode174=13 \def ®{\approx}          
\catcode175=13 \def ¯{\Omega}          
\catcode176=13 \def °{\otimes}          
\catcode177=13 \def ±{\ne}                 
\catcode178=13 \def ²{\le}                   
\catcode179=13 \def ³{\ge}                  
\catcode180=13 \def ´{\upsilon}          
\catcode181=13 \def µ{\mu}                
\catcode182=13 \def ¶{\delta}             
\catcode183=13 \def ·{\epsilon}          
\catcode184=13 \def ¸{\Pi}                  
\catcode185=13 \def ¹{\pi}                  
\catcode186=13 \def º{\beta}               
\catcode187=13 \def »{\partial}           
\catcode188=13 \def ¼{\nobreak\ }       
\catcode189=13 \def ½{\zeta}               
\catcode190=13 \def ¾{\sim}                 
\catcode191=13 \def ¿{\omega}           
\catcode192=13 \def À{\dt}                     
\catcode193=13 \def Á{\gets}                
\catcode194=13 \def Â{\lambda}           
\catcode195=13 \def Ã{\nu}                   
\catcode196=13 \def Ä{\phi}                  
\catcode197=13 \def Å{\xi}                     
\catcode198=13 \def Æ{\psi}                  
\catcode199=13 \def Ç{\int}                    
\catcode200=13 \def È{\oint}                 
\catcode201=13 \def É{\relax\ifmmode\cdot\else\vol\fi}    
\catcode202=13 \def Ê{\relax\ifmmode\,\else\thinspace\fi}
\catcode203=13 \def Ë{\`A}                      
\catcode204=13 \def Ì{\~A}                      
\catcode205=13 \def Í{\~O}                      
\catcode206=13 \def Î{\Theta}              
\catcode207=13 \def Ï{\theta}               
\catcode208=13 \def Ð{\relax\ifmmode\bar\else\expandafter\=\fi}
\catcode209=13 \def Ñ{\overline}             
\catcode210=13 \def Ò{\langle}               
\catcode211=13 \def Ó{\relax\ifmmode\{\else\ital\fi}      
\catcode212=13 \def Ô{\rangle}               
\catcode213=13 \def Õ{\}}                        
\catcode214=13 \def Ö{\sla}                      
\catcode215=13 \def ×{\relax\ifmmode\check\else\expandafter\v\fi}
\catcode216=13 \def Ø{\"y}                     
\catcode217=13 \def Ù{\"Y}  		    
\catcode218=13 \def Ú{\Leftarrow}       
\catcode219=13 \def Û{\Leftrightarrow}       
\catcode220=13 \def Ü{\relax\ifmmode\Rightarrow\else\sect\fi}
\catcode221=13 \def Ý{\sum}                  
\catcode222=13 \def Þ{\prod}                 
\catcode223=13 \def ß{\widehat}              
\catcode224=13 \def à{\pm}                     
\catcode225=13 \def á{\nabla}                
\catcode226=13 \def â{\quad}                 
\catcode227=13 \def ã{\in}               	
\catcode228=13 \def ä{\star}      	      
\catcode229=13 \def å{\sqrt}                   
\catcode230=13 \def æ{\^E}			
\catcode231=13 \def ç{\Upsilon}              
\catcode232=13 \def è{\"E}    	   	 
\catcode233=13 \def é{\`E}               	  
\catcode234=13 \def ê{\Sigma}                
\catcode235=13 \def ë{\Delta}                 
\catcode236=13 \def ì{\Phi}                     
\catcode237=13 \def í{\`I}        		   
\catcode238=13 \def î{\iota}        	     
\catcode239=13 \def ï{\Psi}                     
\catcode240=13 \def ð{\times}                  
\catcode241=13 \def ñ{\Lambda}             
\catcode242=13 \def ò{\cdots}                
\catcode243=13 \def ó{\^U}			
\catcode244=13 \def ô{\`U}    	              
\catcode245=13 \def õ{\bo}                       
\catcode246=13 \def ö{\relax\ifmmode\hat\else\expandafter\^\fi}
\catcode247=13 \def÷{\relax\ifmmode\tilde\else\expandafter\~\fi}
\catcode248=13 \def ø{\ll}                         
\catcode249=13 \def ù{\gg}                       
\catcode250=13 \def ú{\eta}                      
\catcode251=13 \def û{\kappa}                  
\catcode252=13 \def ü{\half}     		 
\catcode253=13 \def ý{\Gamma} 		
\catcode254=13 \def þ{\Xi}   			
\catcode255=13 \def ÿ{\relax\ifmmode{}^{\dagger}{}\else\dag\fi}


\def\ital#1Õ{{\it#1\/}}	     
\def\un#1{\relax\ifmmode\underline#1\else $\underline{\hbox{#1}}$
	\relax\fi}

\def\ron#1#2{{\buildrel{#1}\over{#2}}}	
\def\tdt#1{\on{\hbox{\bf .\kern-1pt.\kern-1pt.}}#1}   
\def\({\eqno(}
\def\li{\eqalignno}
\def\refs{\sect{REFERENCES}\par\medskip \frenchspacing 
	\parskip=0pt \refrm \baselineskip=1.23em plus 1pt
	\def\ital##1Õ{{\refit##1\/}}}


\def\õ#1{
	\screwcount\num
	\num=1
	\screwdimen\downsy
	\downsy=-1.5ex
	\mkern-3.5mu
	õ
	\loop
	\ifnum\num<#1
	\llap{\raise\num\downsy\hbox{$õ$}}
	\advance\num by1
	\repeat}
\def\upõ#1#2{\screwcount\numup
	\numup=#1
	\advance\numup by-1
	\screwdimen\upsy
	\upsy=.75ex
	\mkern3.5mu
	\raise\numup\upsy\hbox{$#2$}}


\catcode`\|=\active \catcode`\<=\active \catcode`\>=\active 
\def|{\relax\ifmmode\delimiter"026A30C \else$\mathchar"026A$\fi}
\def<{\relax\ifmmode\mathchar"313C \else$\mathchar"313C$\fi}
\def>{\relax\ifmmode\mathchar"313E \else$\mathchar"313E$\fi}


\def\ddt#1{\ron{\hbox{\bf .}}{#1}}
\paper

THE SELF-DUAL SECTOR\\ OF QCD AMPLITUDES

\author{G. Chalmers and W. Siegel\footnote{${}^1$}{
 Internet addresses: chalmers and siegel@insti.physics.sunysb.edu.}}\ITP

96-29

June 11, 1996

We provide an action for self-dual Yang-Mills theory which is a simple 
truncation of the usual Yang-Mills action. Only vertices that violate 
helicity conservation maximally are included. One-loop amplitudes in the
self-dual theory then follow as a subset of the Yang-Mills ones. In
light-cone gauges this action is almost identical to previously proposed
actions, but in this formulation the vanishing of all higher-loop amplitudes
is obvious; the explicit perturbative S-matrix is known. Similar results
apply to gravity.

Ü1. Introduction

Certain S-matrix amplitudes in the high-energy, or massless, limit of
quantum chromodynamics take a particularly simple form both at tree and
one-loop level. These amplitudes describe processes where (almost) all
external out-going lines possess the same helicity. Such single-helicity
configurations have a natural interpretation in terms of the scattering of
a self-dual gauge field.

Specifically, the $n$-point gluon tree amplitudes with all, or all but one, 
helicities the same vanish (as implied by a supersymmetric identity [1,2]).
Those with two helicities opposite, the Parke-Taylor amplitudes, have the
momentum dependence [3]
$$ A_n^{\rm tree} (g_1^-,g_2^-, g_3^+, \ldots, g_n^+) = i {Ò12Ô^4\over
Ò12ÔÒ23Ô...Òn1Ô} \ . 
\(1) $$
 We have written the result in a color-ordered form and used the twistor
language [4], also known as spinor helicity [5], to express the helicities
and on-shell massless momenta. All quantities are written in terms of
two-component SL(2,C) (Weyl) spinors; in both matrix and (van der
Waerden) index notation we have for 
$p^2=k^2=0$,
$$ p = |pÔ[p|âÛâp_{ŒÀº} = p_Œ p_{Àº} 
\(2a) $$
$$ ÒpkÔ = p^Œ k_Œ,â[pk] = p^{ÀŒ}k_{ÀŒ} 
\(2b)$$
 where a four-vector is represented as a 2$ð$2 matrix (whose
determinant is the usual Lorentz square). The amplitude with the opposite
helicity configuration is found by complex conjugation. 

Furthermore, the leading-color component of the $n$-point one-loop
gluon amplitudes with all helicities the same has the simple dependence
[6]
$$ A_{n;1}^{[1]} (g_1^+, g_2^+, \ldots, g_n^+) = \sum_{ijkl¼cyclic} -{i\over
192¹^2}{[ij]ÒjkÔ[kl]ÒliÔ\over Ò12ÔÒ23Ô...Òn1Ô} \ .
\(3) $$
 where the sum is over cyclic orderings of any four numbers $i,j,k,l$ in
the range 1 to $n$.  The non-leading-color component is a sum of
permutations of the  leading term [7]. (We refer the reader to [2,8] for a
detailed discussion on the techniques used in calculating tree and loop
amplitudes in gauge theories.) In a supersymmetric theory the
corresponding gluon amplitude vanishes to all orders in perturbation
theory. The loop amplitudes (3) have only two-particle poles and no cuts,
and thus resemble tree graphs. The absence of cuts is due to the
vanishing of the on-shell maximally helicity violating (MHV) tree
amplitudes appearing in the Cutkosky rules. No higher-loop amplitudes in
pure Yang-Mills theory have these simple features; the cuts of two- or
more-loop amplitudes are proportional to phase space integrals of
non-vanishing lower-order amplitudes. 

Bardeen [9] proposed that the simple form of these amplitudes could be
derived from a self-dual Yang-Mills theory. Previously one of us had
pointed out [10] that the light-cone [11] superspace action for self-dual
$N=4$ super Yang-Mills theory is a truncation of the corresponding
non-self-dual action [12] to chiral terms, and had given a Lorentz
covariant component action that generates it. In this paper we show that
the self-dual theory based on the chiral truncation gives the subset of the
Yang-Mills light-cone vertices that are maximally helicity violating. The
S-matrices derived in our formulation of self-dual Yang-Mills theory are
automatically the subset of those in light-cone Yang-Mills theory
consisting of amplitudes of $1-l$ gluons with helicity
$-1$ and all the rest $+1$, where $l$ is the number of loops. Explicitly,
they consist of (1) the tree graphs with one helicity $-1$ and all the rest
$+1$, (2) the one-loop graphs with all helicities $+1$, and (3) no graphs at
all at two or more loops.
 
The two physical polarizations of the gauge field in the light-cone action
are represented in our formulation by the highest and lowest components
of a chiral superfield, as defined in the $N=4$ light-cone supersymmetric
action given below or its N=0,1,2 truncations. Both fields appear in the
theory's truncation to self-dual form. However, the self-dual action given
here is not identical to previous self-dual actions [11,13,14], which have
the field content of only one of the two physical polarizations.
Specifically, as required by Lorentz covariance, the light-cone Yang-Mills
field has two transverse components describing the two helicities, which
are present in {\it both} the self-dual and non-self-dual theories. In the
self-dual theory one of the two components appears only linearly in the
classical action, and thus to order $1-l$ in perturbation theory.

In the following section we derive the self-dual action by truncation of
the usual Yang-Mills action in the light-cone formalism. We prove this 
action describes self-duality and that the truncation preserves Lorentz
covariance, by deriving it from a Lorentz-covariant self-dual form in a
way that is exact within perturbation theory. In section 3 we compare our
action with other actions proposed to describe self-dual Yang-Mills
theory. The other actions, unlike ours, are not Lorentz covariant, have a 
dimensionful coupling constant, and at more than one loop generate 
diagrams that do not relate to Yang-Mills theory. Finally, in section 4 we
speculate on relations to anomalies and string theory.

Ü2. N=4 supersymmetry and self-dual lagrangians

We first consider the light-cone action for $N=4$ supersymmetric 
Yang-Mills theory [12]; the reduction to pure Yang-Mills theory is
achieved by simply dropping the lower-spin fields.

We adopt the notation of [15], so that all quantities are written in terms
of SL(2,C) two-component spinor indices. Four-vectors are written as
$x^{ŒÀŒ}$, and the component $x^{-¬-}$ represents the ``time" coordinate
of the light-cone formalism. Spinor indices are raised and lowered
according to $^à =¦ i_¦$, $^{¬à} =à i_{¬¦}$, and the Lorentz inner
product is $p^2 =-{\rm det}¼p_{ŒÀº}$ .

In the light-cone formalism the field content of the $N=4$ vector
multiplet is described by a complex chiral superfield whose components
contain only the physical states. The chiral superfields relevant to
$N=4$ light-cone superspace are defined by the chirality condition
$$ ÐD^a Ä = 0âÜâÄ(x,Ï,ÐÏ) = exp(Ï^a ÐÏ_a i»_{+¬+})öÄ(x,Ï) \, 
\(4) $$
 in terms of the anticommuting derivatives 
$$ D_a = {»\over » Ï^a} +ÐÏ_a i»_{+¬+}, 
 âÐD^a = {»\over » ÐÏ_a} +Ï^a i»_{+¬+} \ . 
\(5) $$
 Here $a$ is a four-valued index of the internal SU(4) symmetry of N=4
supersymmetry, and we adopt the normalization 
$Çd^4 ϼ Ï^4=1$. In addition, we impose the ``reality" condition on $Ä$, 
$$ D^4 Ä = (i»_{+¬+})^2 ÐÄ \ . 
\(6) $$
 Expanding $Ä$ in $Ï^a$ gives the various component fields, but only
those corresponding to physical polarizations. In $N=4$ light-cone
superspace, $Ä$ and $d^4 Ï$ have helicity assignments 1,-2 respectively
(and opposite for the conjugates). The $Ï$ expansion of $Ä$ is an
expansion in the component fields of helicity equal to 1 minus half the 
order in $Ï$; there are 1,4,6,4,1 fields possessing helicity 
+1,+1/2,0,-1/2,-1. 

The $N=4$ light-cone action can be written simply in light-cone 
superspace [12] as 
$$ S = S_2 + S_{3,c} + S_{3,Ðc} + S_4 
\(7a) $$
$$ S_2 = \f1{g^2}{\rm Tr} Çd^4 x d^4 ϼüÄõÄ 
\(7b) $$
$$ S_{3,c} = \f1{g^2} {\rm Tr} Çd^4 x d^4 ϼ\f13 iÄ(»_+{}^{ÀŒ} Ä)(»_{+ÀŒ}Ä) 
\(7c) $$
$$ S_4 = \f1{g^2} {\rm Tr} Çd^4 x d^4 Ï d^4 ÐÏ ¼\Bigl(\f18[Ä,ÐÄ]^2 
	-\f14[Ä,»_{+¬+}Ä](»_{+¬+})^{-2}[ÐÄ,»_{+¬+}ÐÄ] \Bigr) 
\(7d) $$
 where $S_{3,Ðc}$ is the complex conjugate of $S_{3,c}$. (Note that
$»_+{}^{ÀŒ} ļ»_{+ÀŒ}Ä =-»_{+ÀŒ}ļ»_+{}^{ÀŒ} Ä$.) Further, 
$S_2$ is real because $Ä$ satisfies the reality condition (6). Using this
constraint the action may be written with only $d^4 Ï d^4 ÐÏ$ in a way
where reality is manifest. We have further written $Ä$ in matrix notation
with Hermitian group generators. 

The usual transverse components $A_{ଦ}$ of the gauge fields appear in
$Ä$ as 
$$ Ä={1\over »_{+¬+}} A_{-¬+} +\ldots - Ï^4 »_{+¬+} A_{+¬-} \ . 
\(8)$$
 The two circular polarizations of the gauge fields then reduce to particle
and antiparticle assignments of the complex field $A_{+¬-}$. 

The total helicity of the external fields at any vertex in the action (7)
follows from counting the powers of $Ä$ and $Ï$: $S_2$ and $S_4$ have
total helicity 0, $S_{3,c}$ has +1, and $S_{3,Ðc}$ has $-1$. (Since total
angular momentum is conserved, the helicity may alternatively be read 
off from the spacetime derivatives, which give the orbital angular
momentum.) The vertex which gives the maximal helicity violation is
$S_{3,c}$, while $S_{3,Ðc}$ gives the minimal (negative) violation. Consider
the truncation to $S_2$ and $S_{3,c}$:
$$ S = \f1{g^2} {\rm Tr} Çd^4 x d^4 ϼüÄõÄ +\f13iÄ(»_+{}^{ÀŒ}Ä)(»_{+ÀŒ}Ä). 
\(9) $$
 The $Ï$ expansion generates all of the (3-point) couplings between the
$N=4$ matter fields in which the total out-going helicity is one; it
generates Feynman diagrams, and amplitudes, possessing maximal
helicity violation when regarded as a subset of the complete Lagrangian
(7). In the supersymmetric form (9), we may replace $Ä$ by $öÄ$ since 
there is no $ÐÏ$ dependence.

Upon further reduction to just the non-supersymmetric Yang-Mills fields,
the action (9) becomes 
$$ S = \f1{g^2} {\rm Tr} Çd^4 x ¼Ä_-[õÄ_+ +i(»_+{}^{ÀŒ}Ä_+)(»_{+ÀŒ}Ä_+)] \ .
\(10) $$ 
 We have written the fields as they naturally appear in the $Ï$ expansion
of $Ä$: $Ä_+$ is the lowest component and $Ä_-$ is the highest. This
results in a Jacobian factor of one in going from
$A_{-¬+}$ to $Ä_+$ and $A_{+¬-}$ to $Ä_-$, where the complex fields are
formally treated as independent. Since $Ä_-$ appears only linearly in both
terms in (10), it can be used to count loops; the number of external
$Ä_-$ lines is just $1-l$. (It can absorb the factor $1/\hbar$ multiplying 
the action in the functional integral, just like the dilaton in string 
theory.) Note that the action (10) does not require a dimensionful 
coupling constant; $Ä_-$ and $Ä_+$ have mass dimensions $0$ and $2$. 

Thus, the action (10) is unable to generate diagrams with external $Ä_-$
lines except at tree level, in which case only one external $Ä_-$ state is
possible. The one-loop contributions generate the amplitudes (2), as seen
upon comparison with the pure Yang-Mills sector of the non-self-dual
light-cone theory (7). The other vertices in the YM action (7) are quadratic
in $Ä_-$, and thus generate contributions to S-matrices with more
external lines of helicity $-1$ (i.e., amplitudes that are not MHV). There
are no further loop corrections to the S-matrix from the action (10). 
Furthermore, as we will prove below, this action can be obtained by
quantization of an action that describes self-dual Yang-Mills theory in a
manifestly Lorentz covariant way. 

(The MHV gluon amplitudes calculated in the supersymmetric theory (9)
vanish to all orders in perturbation theory [1,2]. The S-matrix of external
gauge bosons is trivial in this case, although there are non-vanishing
contributions to amplitudes between lower-spin fields.) 

Self-dual Yang-Mills theory is defined only in four spacetime dimensions,
and because of reality properties, only with an even number of time
dimensions. If we include spinors (twistors or physical fermions), then
only 2+2 dimensions is allowed because 4+0 has no Majorana
spinors. (This is also the case relevant to the $N=2$ string [16].) 
However, we are interested in using the self-dual theory to describe a
sector of the physical (non-self-dual) theory, which resides in 3+1
dimensions. We now briefly clarify the differences between the actions
(9-10) in spacetimes with these two signatures. In 3+1 dimensions the
fields $Ä$ and $ÐÄ$ are treated asymmetrically --- they are complex
conjugates, as are $Ï^a$ and $ÐÏ_a$, while $A_{ŒÀº}$ and $x^{ŒÀº}$ are
Hermitian matrices. In this case, the two truncated actions (9-10) are
then complex. 

Alternatively, one can treat our actions in $D=2+2$ dimensions after a
Wick rotation. In this case all covering groups for (super-)space-time
symmetries become real. In particular, the SL(2,C) Lorentz symmetry
becomes SL(2)$°$SL(2), and the internal SU(4) goes into SL(4).
(Furthermore, conformal SU(2,2) $\rightarrow$ SL(4) and superconformal
SU(2,2|4) $\rightarrow$ SL(4|4).) Thus, all the objects
$Ä,ÐÄ,A^{ŒÀº},x^{ŒÀº},Ï^a,ÐÏ_a$ become separately real; $Ï$ and
$ÐÏ$ are then independent, while the constraint (6) determines $ÐÄ$ in
terms of $Ä$.

We complete our discussion of the light-cone self-dual actions in (9-10) by
giving a manifestly Lorentz covariant theory which reproduces them upon
going to the light-cone. We start with the N=4 supersymmetric self-dual
action [10]
$$ S = \f1{g^2} {\rm Tr} Çd^4 x¼üG^{Œº}F_{Œº} +^{aŒ}á_Œ{}^{Àº}_{aÀº}
	+·^{abcd}(\f18 Ä_{ab}õÄ_{cd} +\f14 Ä_{ab}_c{}^{ÀŒ}_{dÀŒ}) \ .
\(11) $$
 The field $G^{μ}$ is an anti-self-dual Lagrange multiplier (which has
mass dimension $2$); the anti-self-dual part of the Yang-Mills field
strength is
$$ F_{Œº} = »_{(Œ}{}^{À©}A_{º)À©} +i [A_{Œ}{}^{À©},A_{ºÀ©}] 
\(12)$$
 All the components are related by N=4 supersymmetry; when truncated
to N$²$2 super Yang-Mills theories, the fields form two separate
multiplets. 

The action can be reduced to the light-cone form at the quantum level. 
We only examine here the non-supersymmetric gauge sector, 
$S = \f1{g^2} {\rm Tr} Çd^4 x¼üG^{Œº}F_{Œº}$. The equations of motion
are 
$$ F_{Œº}=0,ââá^{ŒÀŒ} G_{Œº} =0 \ , 
\(13)$$
 which classically choose only the self-dual part $F_{ÀŒÀº}$ of the
Yang-Mills field strength to survive, while giving the anti-self-dual field
$G_{μ}$ the same field equation that would be satisified by $F_{μ}$ in
the non-self-dual theory. The various Lorentz components expanded out
give 
$$ L=üG^{Œº}F_{Œº} = -üG_{--}F_{++} +G_{+-}F_{+-} -üG_{++}F_{--}
 \(14) $$
 where explicitly
$$ F_{++} = -2i(»_{+¬+}A_{+¬-} -»_{+¬-}A_{+¬+}) +2 [A_{+¬+},A_{+¬-}] 
\(15a)$$ 
$$ F_{+-} = -i(»_{+¬+}A_{-¬-} -»_{+¬-}A_{-¬+} +»_{-¬+}A_{+¬-} -»_{-¬-}A_{+¬+})
	+([A_{+¬+},A_{-¬-}] +[A_{-¬+},A_{+¬-}]) 
\(15b) $$ 
$$ F_{--} = -2i(»_{-¬+}A_{-¬-} -»_{-¬-}A_{-¬+}) +2[A_{-¬+},A_{-¬-}] 
\(15c) $$
 We first choose the light-cone gauge $A_{+¬+}=0$; as usual, the
Faddeev-Popov ghosts decouple. In this gauge the $G_{--}$ term has only
an Abelian component, 
$$ L^{lc}_{--} = i G_{--} »_{+¬+}A_{+¬-} \ ,
\(16) $$
 and may also be functionally integrated out; this enforces
$A_{+¬-}=0$. (The constant Jacobian ${\rm det} ¼»_{+¬+}$ decouples, as in the
Faddeev-Popov determinant of the previous step.) The surviving
contribution for $G_{+-}$ is now also Abelian,
$$ L^{lc}_{+-}= -i G_{+-} (»_{+¬+}A_{-¬-} -»_{+¬-}A_{-¬+} ) \ ,
\(17) $$ 
 and can be solved to give the final expression for the gauge potentials
$$ A_{+ÀŒ} = 0,âA_{-ÀŒ} = »_{+ÀŒ}Ä_+ \ . 
\(18) $$
 We are left with the $L^{lc}_{++}$ term; upon relabelling 
$G_{++}=i\phi_-$ we find the action (9). 

The manipulations we have just performed are exact within perturbation
theory, and prove the equality of the covariant (11) and light-cone (10)
forms of the S-matrix elements to all orders, in the gauge sector. The
complex-conjugate Lagrangian may be derived using an (anti-) self-dual
covariant action, i.e., with dotted and undotted indices reversed in (11).
(As usual, we freely invert the ``spatial" derivative $»_{+¬+}$, which is legal
with appropriate boundary conditions. Also, since the theory is Lorentz
covariant, $»_{+¬+}^{-1}$ cannot generate poles by itself in $D>2$. 
Furthermore, since we neglect only determinants of free derivatives, any
modes which might be missed by inverting such derivatives are those that
decouple.) 

Finally, we make a few remarks about how helicity is defined and its
relation in the self-dual and non-self-dual actions. The simplest way to
define helicity is in terms of field strengths. This method is not only
Lorentz and gauge covariant, but also applies to interacting states. For
example, $F_{ÀŒÀº}$ describes helicity +1, while $F_{Œº}$ (or $G_{Œº}$ in the
self-dual formulation, where $F_{μ}=0$) describes $-1$. The helicity is
simply half the number of dotted minus undotted indices, which follows
from the fact that any field strength satisfies a Weyl equation on each
spinor index. This translates into counting the twistors that carry these
indices: In the free theory, or for asymptotic states,
$$ F_{ÀŒÀº} = p_{ÀŒ}p_{Àº}f_+,âF_{Œº}¼{\rm or}¼G_{Œº} = p_Œ p_º f_- 
\(19)$$
 in terms of some scalar twistor-space functions $f_à$. These expressions
have close analogs in ordinary coordinate (or momentum) space; in the
usual Yang-Mills theory in the light-cone gauge, where
$A_{+¬+}=0$ and $A_{-¬-}$ is eliminated by its field equation, we have
$$ F_{ÀŒÀº} = -i»_{+ÀŒ}»_{+Àº}»_{+¬+}{}^{-1}A_{-¬+} +O(A^2),â
	F_{Œº} = -i»_{Œ¬+}»_{º¬+}»_{+¬+}{}^{-1}A_{+¬-} +O(A^2) 
\(20)$$
 on shell. In the LMP-type light-cone gauge for self-dual Yang-Mills we
have 
$$ F_{ÀŒÀº} = -i»_{+ÀŒ}»_{+Àº}Ä_+,âF_{Œº} = 0 
\(21)$$
$$ G_{Œº} = »_{+¬+}{}^{-1}á_{Œ¬+}»_{+¬+}{}^{-1}á_{º¬+}Ä_- 
	= »_{Œ¬+}»_{º¬+}»_{+¬+}{}^{-2}Ä_- + O(Ä^2) 
\(22) $$
 In 2+2 dimensions, we have the freedom to scale $p_Œ$ and $p_{ÀŒ}$
oppositely in $p_{ŒÀº}=p_Œ p_{Àº}$. (In 3+1 the invariance is a phase, and
we generally have to write $p_{ŒÀº}=àp_Œ p_{Àº}$ to treat both positive
and negative energy. These problems are also avoided by our Wick
rotation from 2+2.) This allows us to choose
$$ p_+ = 1âÜâp_{ÀŒ} = p_{+ÀŒ} \ .
\(23)$$
 This makes Feynman graph calculations in the self-dual theory almost
indistinguishable from twistor calculations, since noncovariant vertex
factors $p_{+ÀŒ}$ can be replaced with covariant twistors $p_{ÀŒ}$ after
being expressed in terms of (on-shell) external momenta.

Ü3. Relations to other proposed self-dual actions

Except for the $Ï$ integration, the above truncated N=4 light-cone action
(9) is the one proposed by Leznov and Mukhtarov, and Parkes (LMP) [11]
to describe self-dual Yang-Mills theory, 
$$ S_{LMP} = \f1{Â^2} {\rm Tr} Çd^4 x¼üÄõÄ
	+\f13iÄ(»_+{}^{ÀŒ}Ä)(»_{+ÀŒ}Ä). 
\(24) $$
 However, the action we use has several important differences. The most
important is that, after truncation to the non-supersymmetric Yang-Mills
sector, we have ÓtwoÕ polarizations, as required for Lorentz invariance,
not one. In the kinetic term the lowest order in $Ï$ component of the
superfield $Ä$ (helicity +1) couples to the highest one (helicity -1). 
 
The fact that the LMP action has only one field has two immediate
consequences: (1) The LMP action is not Lorentz invariant, not even in a
hidden way. (2) The coupling constant in the LMP action has the wrong
(engineering) dimension. The above N=4 action, and its $Ä_à$ truncation,
have neither of these problems.

We now compare the S-matrices of our action to those of the LMP action
in the non-supersymmetric case. (The supersymmetric forms are almost
trivial since all loop amplitudes vanish for both theories.) (1) In our case
the propagator has a ``+" at one end and a ``$-$" at the other; the vertex
has 2 $+$'s and a $-$. In the LMP case no lines are distinguished. (2) There
is no difference at the tree level, since tree S-matrices vanish, except for
the 3-point vertex, which is non-vanishing in 2+2 dimensions. (In 3+1,
kinematic constraints force it to vanish.) The three-point contribution is
indistinguishable in the two theories because of the symmetry of the
vertex, and because the normalization can be absorbed by a redefinition
of the coupling or of $Ä_-$. (3) At the 1-loop level the LMP action gives
the same result ÓexceptÕ for an additional factor of 1/2, since there is
only one field and not two. As usual for one-loop graphs, this
normalization cannot be modified. (4) At higher loops all graphs vanish for
our action. There is no such implication for the LMP action, which
apparently has higher-loop contributions.

Another action to compare against is that proposed by Donaldson, and Nair
and Schiff [14], based on Yang's [13] form of the self-dual equations
(YDNS). We find a similar action from the above covariant form (11) by
slightly modifying the above steps to the light-cone. As before, we choose
the gauge $A_{+¬+}=0$ in (14) and functionally integrate out
$G_{--}$, so $A_{+¬-}=0$. Instead of examining the $G_{+-}$ term, however,
we Abelianize the $G_{++}$ term by the field redefinitions
$$ A_{-¬-} = -i e^{-iÄ}»_{-¬-}e^{iÄ},âA_{-¬+} = -i 
 e^{-iÄ} (»_{-¬+} +i A'_{-¬+})e^{iÄ},â G_{++} = e^{-iÄ}G'_{++}e^{iÄ} \ . 
\(25)$$
 The $L_{++}$ term is then 
$$ L_{++} = - G'_{++} »_{-¬-}A'_{-¬+} \ . 
\(26) $$
 Integrating out $G'_{++}$ sets $A'_{-¬+} =0 $, after dropping the irrelevant
Jacobian factor ${\rm det}¼»_{-¬-}$. 

Up till now all Jacobians have been constants.  Another type of trivial
Jacobian is one of a functional determinant involving no derivatives:  If
such determinants are written in terms of Faddeev-Popov-like ghosts,
the ghosts have nonderivative propagators.  Such determinants produce
$¶^4(0)$ terms, which can be neglected.  (For example, they vanish in
dimensional regularization.)  The Jacobian from the change of variables
(25) reduces to that for the first redefinition, times nonderivative
determinants of this type.  This remaining contribution to the effective
action can be represented by a Faddeev-Popov-like expression
$$ S_c = {\rm Tr} Çd^4 x¼÷C Q (e^{-iÄ}»_{-¬-}e^{iÄ})â{\rm with}â QÄ = C \ , 
\(27) $$
 where $Q$ is a BRST-like operator, which acts in the same way as a
derivative or variation. The action may be re-organized as 
$$ S_c = -{\rm Tr} Çd^4 x¼ (e^{-iÄ} ÷C e^{iÄ}) »_{-¬-} ( e^{-iÄ} Qe^{iÄ}) \ . 
\(28) $$
 We next perform two successive field redefinitions on the ghosts, the
Jacobians of which are trivial ($¶^4(0)$ and constant terms,
respectively), 
$$ ÷C = e^{iÄ} ÷C' e^{-iÄ},ââ÷C' = {1 \over »_{-¬-}}÷C''\ , 
\(29) $$
 and obtain the contribution
$$ S_c = {\rm Tr} Çd^4 x¼ ÷C'' ( e^{-iÄ} Q e^{iÄ}) \ . 
\(30) $$
 This ghost term may be path-integrated out since it is algebraic. The
final expression for the potential is
$$ A_{+ÀŒ} = 0,âA_{-ÀŒ} = -i e^{-iÄ}»_{-ÀŒ}e^{iÄ}. 
\(31) $$
 The resulting action comes from the $ L^{lc}_{+-}$ term, and gives the
Yang field equation, but from a two-field action
$$ S = -i¼{\rm Tr} Çd^4 x ¼G_{+-} »_+{}^{ÀŒ} (e^{-iÄ}»_{-ÀŒ}e^{iÄ})
 \(32) $$
 This action thus also gives S-matrices equal to those of non-self-dual
Yang-Mills theory restricted to certain helicities.

On the other hand, the YDNS action gives S-matrices that disagree in the
same way as described above for the LMP action.  The YDNS action gives
the same field equations as (32), but in terms of one field instead of
two:
$$ ¶S = Çf(Ä)¶G_{+-} +h(Ä,G_{+-})ëÄ,â¶S_{YDNS} = Çf(Ä)ëÄ;
	âëÄ ­ -ie^{-iÄ}¶e^{iÄ}; 
\(33) $$
 where we have used the covariant variation $ëÄ$.  (Using the
covariant variation instead of the naive one just introduces another
trivial determinant.)  The one-loop S-matrix is expressed in terms of the
one-loop effective action, which is the determinant of the second
functional derivative of the classical action:
$$ S_{eff,YDNS} = -ü ln¼det\left( {¶^2 S_{YDNS}\over ëÄëÄ} \right)
	=  -ü ln¼det\left( {¶f\over ëÄ} \right) $$
$$ \li{ S_{eff} & = -ü ln¼det\pmatrix{ \displaystyle{¶^2 S\over ëÄëÄ} &
	\displaystyle{¶^2 S\over ëĶG_{+-}} \cr & \cr
	\displaystyle{¶^2 S\over ¶G_{+-}ëÄ} &
	\displaystyle{¶^2 S\over ¶G_{+-}¶G_{+-}}} \cr \strut & \cr
	& = -ü ln¼det\pmatrix{ \displaystyle{¶h\over ëÄ} &
	\displaystyle{¶f\over ëÄ} \cr & \cr \displaystyle{¶f\over ëÄ} & 0}
	=  -ln¼det\left( {¶f\over ëÄ} \right) & (34) \cr}  $$
 We have thus proven the equivalence of our modifications of the LMP and
YDNS actions, and that the original LMP and YDNS actions give the same
one-loop S-matrices (both differing from ours by a factor of 1/2).

The YDNS action has also been proposed to describe the N=2 (open) string
[16]. However, it is also possible to interpret that string in terms of our
two-field modification of that action: States in that string in different
pictures are usually interpreted as the same state, since their couplings
are the same. However, in ordinary QCD we know maximally helicity
violating couplings are helicity independent. If we use helicity (i.e.,
Lorentz transformations) to distinguish otherwise-identical states [17],
then (at least) two different states appear in Lorentz invariant
amplitudes.

Similar remarks can be made regarding gravity. The analog of the YDNS
action for self-dual gravity, the Pleba«nski action [18], must be modified to
contain the fields describing both $à2$ helicities. The light-cone action
for gravity [20] can easily be truncated for maximal helicity violation to
give the analog of the LMP action [19]; the infinite number of terms
reduce to one interaction plus the kinetic term. All the other terms
generate amplitudes which contain at least one more negative-helicity 
external state. 

Remarks made in the introduction carry over to the gravitational case. As
with Yang-Mills theory, the MHV graviton scattering amplitudes vanish at
tree level and must be cut-free at one loop. However, the all-plus
one-loop scattering amplitudes have not been calculated beyond
four-point [21]; complete solutions to the self-dual theory, unlike SDYM,
are not known explicitly. 

Ü4. Discussion

Bardeen has conjectured that these amplitudes are related to anomalies. 
The effective action for our self-dual theory receives contributions only
at one loop. A possible candidate for this one-loop contribution is the
trace anomaly, which leads to very simple effective actions in
two-dimensional theories. For  example, in the Schwinger model a fermion
loop generates exactly $Fõ^{-1}F$ for the effective action. The 4D analog
would be $Fõ^·F/·$ $=$ $F^2/· +F(\ln¼õ)F$, where the divergent term
vanishes upon integration for self-dual $F$ (and $õ$ is gauge covariant).
We have been unable to verify, however, that the latter term is in fact
the complete effective action. 

Another, more interesting, possibility is that the one-loop contribution
might be generated by a ÓlocalÕ term in the effective action through the
introduction of extra fields. This also has an analog in the Schwinger
model, where the fermion's contribution to 
$S_{\rm eff}[A]$ may be reproduced by introducing an extra scalar field
(the fermion-anti-fermion condensate that comes from bosonization),
resulting in the Stueckelberg action for a massive vector.

The existence of a local term is suggested not only by the appearance of
only poles in the one-loop S-matrices, but by string theory:  The N=2 open
string is known to describe self-dual Yang-Mills theory [12] (or its
supersymmetric generalizations [15]).  One-loop diagrams in open-string
theory are equivalent to ÓtreeÕ graphs in the combined theory of open and
closed strings [22]. In the one-loop planar graph, the loop can be pulled
out to represent a closed string propagator connecting an open string tree
to the vacuum; the one-loop double-twisted graph can be stretched to
produce a closed string propagator connecting two open string trees. This
suggests the introduction of fields without physical polarizations to
represent the closed string. A likely candidate would be a dilaton, namely
the Weyl scale mode of the metric, which in ordinary gravity has no
physical degrees of freedom (although it has a nontrivial kinetic term).
Also, it couples to the trace of the energy-momentum tensor, which
relates to the previous conjecture concerning the trace anomaly.

Explicit calculations in string theory [23], however, have indicated the
vanishing of all one-loop graphs with more than three external lines in all
N=2 string theories. These string results are in direct contradiction with
field theory. This suggests some subtlety was missed, possibly signalling
the presence of an anomaly in the worldsheet theory describing the
string. 

ÜNOTE ADDED

After this work was completed, Cangemi [24] showed by explicit
calculation that the light-cone action for self-dual Yang-Mills theory
gives the one-loop S-matrices for ordinary Yang-Mills theory with all
external helicities the same.

ÜACKNOWLEDGMENTS

We thank Zvi Bern for bringing Bardeen's paper to our attention. This work
was supported in part by the National Science Foundation Grant No.¼PHY
9309888.

\refs

£1 M.T. Grisaru, H.N. Pendleton, and P. van Nieuwenhuizen, \PR 15
	(1977) 996; \\ 
	 M.T. Grisaru and H.N. Pendleton, \NP 124 (1977) 333.
 £2 M. Mangano and S.J. Parke, ÓPhys. Rep.Õ É200 (1991) 301.
 £3 S.J. Parke and T. Taylor, \NP 269 (1986) 410, \PR 56 (1986) 2459;\\
	F.A. Berends and W.T. Giele, \NP 306 (1988) 759.
 £4 R. Penrose, ÓJ. Math. Phys.Õ É8 (1967) 345, 
	ÓInt. J. Theor. Phys.Õ É1 (1968) 61;\\
	M.A.H. MacCallum and R. Penrose, ÓPhys. Rep.Õ É6C (1973) 241;\\
	A. Ferber, \NP 132 (1978) 55.
 £5 P. De Causmaecker, R. Gastmans, W. Troost, and T.T. Wu, 
	\NP 206 (1982) 53;\\
	F.A. Berends, R. Kleiss, P. De Causmaecker, R. Gastmans, W. Troost,
	and T.T. Wu, \NP 206 (1982) 61;\\
	Z. Xu, D.-H. Zhang, and L. Chang, \NP 291 (1987) 392;\\
	J.F. Gunion and Z. Kunszt, \PL 161 (1985) 333;\\
	R. Kleiss and W.J. Sterling, \NP 262 (1985) 235. 
 £6 Z. Bern, G. Chalmers, L. Dixon, and D.A. Kosower, \PR 72 (1994) 2134,\\
	G.D. Mahlon, \PRD 49 (1994) 4438.
 £7 Z. Bern and D.A. Kosower, \NP 362 (1991) 389, \\ 
	Z. Bern, L. Dixon, D.C. Dunbar, and D.A. Kosower, \NP 425 (1994) 217.
 £8 Z. Bern, L. Dixon, D.A. Kosower, preprint SLAC-PUB-7111 (Feb 1996), 
	hep-ph/9602280.
 £9 W.A. Bardeen, Selfdual Yang-Mills theory, integrability and
	multi-parton amplitudes, preprint FERMILAB-CONF-95-379-T
	(August 1995).
 £10 W. Siegel, \PRD 46 (1992) R3235.
 £11 A.N. Leznov, ÓTheor. Math. Phys.Õ É73 (1988) 1233,\\
	A.N. Leznov and M.A. Mukhtarov, ÓJ. Math. Phys.Õ É28 (1987) 2574;\\
	A. Parkes, \PL 286B (1992) 265.
 £12 S. Mandelstam, \NP 213 (1983) 149;\\
	L. Brink, O. Lindgren, and B.E.W. Nilsson, \NP 212 (1983) 401.
 £13 C.N. Yang, \PR 38 (1977) 1377.
 £14 S. Donaldson, ÓProc. Lond. Math. Soc.Õ É50 (1985) 1,\\
	V.P. Nair and J. Schiff, \PL 246B (1990) 423.
 £15 S.J. Gates, Jr., M.T. Grisaru, M. Ro×cek, and W. Siegel, ÓSuperspaceÕ or 
	 ÓOne thousand and one lessons in supersymmetryÕ
	(Benjamin/Cummings, Reading, 1983).
 £16 H. Ooguri and C. Vafa, ÓMod. Phys. Lett.Õ ÉA5 (1990) 1389,
	\NP 361 (1991) 469, É367 (1991) 83;\\
	N. Marcus, \NP 387 (1992) 263.
 £17 W. Siegel, \PR 69 (1992) 1493.
 £18 J.F. Pleba«nski, ÓJ. Math. Phys.Õ É16 (1975) 2395.
 £19 W. Siegel, \PRD 47 (1993) 2504.
 £20 J. Scherk and J.H. Schwarz, ÓGen. Rel. and Grav.Õ É6 (1975) 517,\\
	M. Kaku, \NP 91 (1975) 99,\\
	M. Goroff and J.H. Schwarz, \PL 127 (1983) 61.
 £21 M.T. Grisaru and J. Zak, \PL 90B (1980) 237,\\
	Z. Bern, D.C. Dunbar, and T. Shimada, \PL 312B (1993) 277,\\
	D.C. Dunbar and P.S. Norridge, \NP 433 (1995) 181.
 £22 D.J. Gross, A. Neveu, J. Scherk, and J.H. Schwarz, \PL 31B (1970)
	592,\\
	C. Lovelace, \PL 34B (1971) 500,\\
	E. Cremmer and J. Scherk, \NP 50 (1972) 222.
 £23 N. Berkovits and C. Vafa , \NP 433 (1995) 123.
 £24 D. Cangemi, Self-dual Yang-Mills theory and one-loop like-helicity
	QCD multi-gluon amplitudes, preprint UCLA/96/TEP/16 (May 1996),
	hep-th/9605211. 

\bye